\documentstyle[amssymb,preprint,aps]{revtex}
\begin{document}
\title{Fermi Surface of the One-dimensional Kondo Lattice Model}
\author{S. Moukouri and L. G. Caron}
\address{Centre de recherche en physique du solide and d\'epartement de physique,\\
Universit\'e de Sherbrooke, Sherbrooke, Qu\'ebec, Canada J1K 2R1}
\date{17 July 1996}
\maketitle

\begin{abstract}
We show a strong indication of the existence of a large Fermi surface in the
one-dimensional Kondo lattice model. The characteristic wave vector of the
model is found to be $k_F=(1+\rho )\pi /2$, $\rho $ being the density of the
conduction electrons. This result is at first obtained for a variant of the
model that includes an antiferromagnetic Heisenberg interaction $J_H$
between the local moments. It is then directly observed in the conventional
Kondo lattice $(J_H=0)$, in the narrow range of Kondo couplings where the
long distance properties of the model are numerically accessible.
\end{abstract}


\section{INTRODUCTION}

Many experiments tell us that one of the low temperature states of the heavy
fermion materials is a Fermi liquid whose quasiparticle masses are $10^2$ to 
$10^3$ larger than those of the normal metals\cite{stewart}. It is commonly
believed that the heavy Fermi liquid state is one of the possible ground
states of the Kondo Lattice model $(KLM)$. The most popular description of
the heavy fermion state is that below a characteristic temperature $T_{coh}$ 
$(T_{coh}\lesssim 10K)$, the conduction electrons ($c-$electrons) and the
local moments ($f-$electrons) have common excitations. This picture leads to
the important fact that the localized spins also participate in the Fermi
surface $(FS)$. Thus, the $FS$ has a large area. De Haas-Van Alphen
measurements\cite{johanson} of some heavy fermion compounds have shown that
the $FS$ has a $f$ character. Ab-initio Local Density approximation $(LDA)$%
\cite{koelling} computations of the band structure of heavy fermions also
predict a large $FS$, although these calculations fail to reproduce heavy
masses. These results are indeed expected from the Luttinger theorem which
states that the volume of the $FS$ is unchanged by the electron-electron
interaction. Intuitively, however, it is not straigthforward to understand
in the framework of the $KLM$, how the $f$-electrons can be included in the $%
FS$ since there is no explicit hybridization between the $c$-electrons and
them. $LDA$ computations have shown the existence of a narrow $f$ bandwith
which is due to hybridization\cite{boring}. The $KLM$ itself is an effective
model of the periodic Anderson lattice $(PAM)$ in the limit of nearly
integral valence and of strong Coulomb correlation. One may thus wonder
wether or not the residual $f$-electrons itinerant character present in the
strong coupling limit of the $PAM$ plays a crucial role in the formation of
the heavy Fermi liquid state.

Analytical\cite{fazekas}\cite{fujimoto} and numerical\cite{tsunetsugu}\cite
{ueda}\cite{moukouri} studies have focused on the $FS$ of the $1D$ $KLM$.
Their results remain very controversial. In the following, we will study the 
$1D$ $KLM$ numerically. A numerical study of the $1D$ $KLM$ presents two
essential difficulties. The first one arises from the very low energy scale
of the Kondo physics which requires the investigation of lattices of very
large sizes. In real materials, the heavy masses involve a very small value
of the quasiparticle weight. For the $1D$ model, in the physical range of
parameters $(J_K\ll 1)$, the size of the expected singularity in the
electron momentum distribution $n_c(k)$ is likely to be very small. The
second problem is the occurrence of a ground-state phase transition from a
paramagnetic $(PM)$ state at weak couplings to a ferromagnetic $(FM)$ state
at strong couplings\cite{tsunetsugu}\cite{moukouri}. Therefore, the results
of the strong coupling regime where the model converges more rapidly to the
thermodynamic limit cannot be extrapolated to the weak-coupling region where
size effects are still significant even in very long chains. We will show
that the study of a $KLM$ (Eq. 1) in which the strong coupling regime is
smoothly connected to the weak coupling one can give insight of the
existence of a large $FS$ for the usual $KLM$. Then, a careful analysis of
the usual $KLM$ will be made. A large $FS$ means that the Fermi wave vector
is located at $k_F=k_{F_c}+\frac \pi 2$, $k_{F_c}$ being the Fermi wave
vector of the $c-$electrons only. We will consider the following $KLM$:

\begin{equation}
H=-t\sum_{is}(c_{is}^{+}c_{i+1s}^{}+h.c.)+J_K\sum_i{\bf S}_{ic}^{}\cdot {\bf %
S}_{if}^{}+J_H\sum_i{\bf S}_{if}^{}\cdot {\bf S}_{i+1f}^{}  \eqnum{1}
\end{equation}

Where ${\bf S}_{ic}^\alpha =\frac 12\sum_{s,s^{\prime }}c_{is}^{+}\sigma
_{ss^{\prime }}^\alpha c_{is^{\prime }}^{}$ and ${\bf S}_{ic}^\alpha $ is a
localized spin. The hopping integral $t$ is set to $1$. A direct Heisenberg
exchange term between local moments $J_H$ is introduced here. We have
recently argued\cite{moukouri} that the occurence of a ferromagnetic phase
transition in the $KLM$ in the strong coupling region is due to the fact
that the $RKKY$ interaction becomes ineffective. One would thus expect that
a sufficiently strong antiferromagnetic Heisenberg coupling between the
local moments can stabilize a $PM$ ground state. In the conventional $KLM$,
this term is usually omitted because typical lattice parameters in the heavy
fermion compound are $3.5-4\AA $ while the ionic radii of the $f$ ions are
less than $1\AA $, so that the overlap between the $f$ orbitals is
negligible. It should be noted that the strong $J_K$ regime of the
Hamiltonian $(1)$ is relevant in the study of High-$T_c$ materials. In this
case, $J_H$ is the superexchange interaction between copper ions. The double
occupancy in the $c-$electron band is naturally suppressed by $J_K$, so that
one need not include an explicit repulsion term between these electrons. We
have investigated Hamiltonian $(1)$ using the density matrix renormalization
group $(DMRG)$ method\cite{white}. We have chosen an algorithm with open
boundary conditions. We keep between $64$ and $150$ states in the two
external blocks. These states are labelled by the $z-$component of the total
spin $S_T^z$. The ground state corresponds to the lowest state with $S_T^z=0$%
. The maximum truncation error is in the order of $10^{-4}$. Although we
have reached $N=60$ sites, the longest distance in the calculation of the
correlation function is $L=22$. Because we have first built lattices of $20$
sites before we start to calculate the correlation functions. This way, we
minimize the end effects and density fluctuations that are larger in the
early steps of the algorithm.

\section{RESULTS FOR THE KONDO-HEISENBERG CASE $(J_H\neq 0)$}

The Hamiltonian $(1)$ has {\it a }${\it PM}${\it \ ground state with a
Luttinger Fermi surface} when $J_H=0.5$ in the strong $J_K$ regime. We have
chosen our value for the Heisenberg coupling on the grounds that it is
necessary that $J_H$ exceeds the effective $FM$ coupling $J_{eff}^{\max }$
between the $f$ electrons to stabilize the $PM$ ground state for all $J_K$.
Our first choice was $J_H=0.1$. For this value, we found that the ground
state is $PM$ in the weak $J_K$ region, $FM$ for intermediate couplings,
then $PM$ for strong $J_K$. We can estimate $J_{eff}$ by using the results
of the strong coupling expansion\cite{sigrist}. $J_{eff}J_K/t^2$ is
approximatively $0.05$, $0.2$ and $0.25$ respectively for the partial
band-fillings $\rho =0.25,0.5$ and $0.75$. Then, knowing that the strong
coupling region starts for $J_K$ of order $1$, one obtains the upper bound, $%
J_{eff}^{\max }\simeq 0.25$ for $t=1$. Our value of $0.5$ thus include a
security factor. In Fig.1, we show $n_c(k)$ (the Fourier transform of $%
\left\langle c_{i\sigma }^{+}c_{j\sigma }^{}\right\rangle $) for the
band-fillings $\rho =0.25,0.5$ and $0.75$ for $J_K=10$. $n_c(k)$ at $0.25$
and $0.75$ are affected by density fluctuations, since the band-filling is
not constant during the $DMRG$ iterations. Nevertheless, clean singularities
are observed at $k_F=0.625\pi $, $0.75\pi $ and $0.875\pi $. The magnetic
structure factor of the localized electrons $S_f(k)$ (the Fourier transform
of $\left\langle S_{if}^zS_{jf}^z\right\rangle $), shown in Fig.2, presents
a maximum at $2k_F=0.75\pi ,0.5\pi $ and $0.25\pi $ respectively. These
values correspond to $2\pi -2k_F$, since $2k_F$ is greater than $\pi $.
Clearly, neither the bare $c-$electron nor the bare $f-$electron signatures
are detected. There are instead unique compound-particles propagating with a
characteristic wave vector at $k_F$. An analogy can be made with the Hubbard
model\cite{lacroix}. At half-filling, in the $J_K=\infty $ limit, all the
conduction electrons form on-site singlets with the localized spins, so that
the overall system is in a singlet state. The non half-filled cases
correspond to the introduction of holes in the system. These holes which can
hop from site to site are associated with the $N-N_c$ unpaired $f-$%
electrons, $N_c$ being the number of $c-$electrons. Obviously, double
occupancy of holes is forbidden : one has a $U=\infty $ Hubbard model of $%
\rho _h=1-\rho $ hole density. Depletion effects\cite{nozieres} are also
observed in Fig.2: the reduction of the conduction electron density
increases the tendency to magnetism. The maximum of $S_f(k)$ increases when $%
\rho $ decreases.

Now we wish to discuss how the system evolves when $J_K$ is reduced. We show
in Fig.3 $n_c(k)$ at $\rho =0.5$ for $J_K=10,8,6,4,3,2.5$ and $2$. The
height of the singularity at $k_F$ decreases as $J_K$ is reduced.
Concurrently the drop at $k_{F_c}$ which, was negligibly small in the strong 
$J_K$ case, increases. The $c-$electron character is progressively enhanced.
The local spin-spin correlation which is $\left\langle
S_{ic}^{}S_{if}^{}\right\rangle =-.3748\simeq -\frac 34\rho $ at $J_K=100$,
is equal to $-.366$ at $J_K=10$ and $-.183$ at $J_K=2$. The deviation of
this quantity from the perfect on-site singlet value $-\frac 34\rho $ means
that the singlet clouds have a spatial extension at lower $J_K$. At $J_K=2$,
short-range effects coexist along with the long-range behavior of the
system. Clearly, it becomes hard to define the exact position of the $FS$.
This can be better illustrated in $S_f(k)$ (Fig.4) at $\rho =0.5$. A new
peak appears at $k=\pi $ at small $J_K$, signaling short-range
antiferromagnetic correlations. At $J_K=1.5$ (not shown here), the peak at $%
2k_F$ is not seen. We can no longer observe the long-distance behavior of
the model because of the finite-size effects (long correlation length; see
the discussion below). Since there is no phase transition in the system, the
weak and the strong $J_K$ regimes are continuously connected. We thus
believe that {\it the }${\it FS}${\it \ is large even at smaller} $J_K$. The
above discussion is similar to the one made by Kotliar in the framework of
the $PAM$\cite{kotliar}. The local singlets of the strong-coupling limit are
obtained as the Kondo resonances are pulled out of the $c-$electron band by
increasing $J_K$. The $FS$ is conserved during this process.

\section{RESULTS FOR THE CONVENTIONAL KLM $(J_H=0)$}

We now discuss if the above conclusions can be extended to the $PM$ phase of
the conventional $KLM$. At first sight, one can argue that the conventional $%
KLM$ is adiabatically reached by taking the limit $J_H\rightarrow 0$. Thus
the conventional $KLM$ may have a large $FS$ in its $PM$ phase. It should be
noted from the above results and from the knowledge of the occurrence of a
phase transition, that the range of $J_K$ in which we can expect to detect a
large $FS$ with numerical methods in the conventional $KLM$ is narrow. An
estimation of this range can be obtained by examining the Kondo coherence
length $\xi _K$ of the one-inpurity problem. $\xi _K=v_F/T_K$, where $v_F$
is the Fermi velocity and $T_K$ the Kondo temperature. The observation of
the long range properties of the model is only possible at distances $r\gg
\xi _K$. For $r\lesssim \xi _K$ finite size effects dominate, only
short-range effects governed by the $RKKY$ interaction will be observed. S\o
rensen and Affleck have recently made an accurate computation of $\xi _K$%
\cite{sorensen}. Some typical values are $\xi _K=1,2,4.85,8$ and $23$ for $%
J_K=2.5,2,1.5,1.25$ and $1$ respectively. Clearly, for $J_K\lesssim 1.25$
the long-range behavior of the model is not attainable since the longest
distance in our study is $L=22$. Moreover, the $FM$ transition occurs at $%
J_K\simeq 1.5$ for $\rho =0.5$ and $J_K\simeq 2.75$ for $\rho =0.75$. At low
band-fillings, depletion effects enhance the $FM$ instability. The $PM$
phase boundary is shifted towards weak couplings where $\xi _K$ becomes very
large. Hence, at quater filling where there is no density fluctuations, this
range is very narrow. Thus we have chosen to study the $KLM$ at $\rho =0.75$
in the range $1.25\lesssim J_K\lesssim 2.5$. In Ref.\cite{moukouri}, we have
found that $n_c(k)$ displays a sharp drop at $k_{F_c}$ and $S_f(k)$ presents
a maximum at $2k_{F_c}$ in the small $J_K$ regime. As $J_K$ was increased,
these features vanished before the phase transition was reached. But we were
unable to draw a firm conclusion about the location of the $FS$. A more
careful analysis will now show that we had observed a short-range effect
governed by the $RKKY$ interaction. Furthermore, we will identify the true
long-range properties of the model which are not easily observable. In
Fig.5, we display $S_f(k)$ at $\rho =0.75$ for $J_K=1.25,1.5,1.75,2,$ and $%
2.5$. Starting from $J_K=1.25$, we can only detect the $RKKY$ maximum at $%
2k_{F_c}=0.75\pi $. For $J_K=1.5$, the maximum of $S_f(k)$ is still located
at the $RKKY$ wave vector. But one can also observe a local maximum at the
position of the large $FS$ at $2k_F=0.25\pi $ . At $J_K=1.75$ and $2$, the
height of the $RKKY$ maximum decreases. At the same time, the height of the
maximum at $2k_F$ increases. This trend is unambigously confirmed at $J_K=2.5
$, where the maximum at $2k_F$ is now the highest. The height of these
maxima however, is still smaller than the one at $2k_{F_c}$ for $J_K=0.5$ in
ref $[9]$, and are thus harder to detect. Finally, $n_c(k)$, shown in Fig.6,
corroborates the existence of the large $FS$ in the $KLM$. $n_c(k)$ drops
monotonously when $k<k_F=0.875\pi $, shows a small plateau just before $k=k_F
$, and then drops abruptly at $k\simeq k_F$. The width of the plateau
shrinks and then becomes undetectable when $J_K$ is decreased. At the same
time, the drop at $k_{F_c}$ is enhanced indicating that the short range
effects are becoming dominant. This is consistent with the results for $%
S_f(k)$.

\section{CONCLUSION}

In summary, we have shown that the $KLM$ with a direct exchange Heisenberg
coupling has a large $FS$ in the strong Kondo coupling limit. We have shown
that this phase is continously connected to the weak Kondo coupling regime.
As a consequence, the latter model has a large $FS$ in the small coupling
regime. The conventional $KLM$ can continously be reached by taking the
limit of vanishing Heisenberg coupling. We have concluded from this that its 
$FS$ should have a large area in its $PM$ phase. Direct numerical
computations made on the $KLM$ support the existence of a large $FS$. 
The height of the singularity in the electron momentum distribution decreases
and becomes very small as the Kondo interaction goes towards the weak
coupling region. 
These features are however very hard to observe. A more accurate 
study is certainly needed to confirm our conclusions.
 Finally, the nature of the
heavy quasiparticles appear to be different from that proposed in the
renormalized band structure studies\cite{cyrot}. In the renormalized band
picture, the heavy quasiparticles result from the small hybridization
between the conduction electrons and the renormalized $f-$bands pinned at
the Fermi level of the conduction electron sea. This 
requires the number of $f$-electrons per
site to be sligthly less than one. Our results suggest that
these are instead loosely bound states made up of conduction electrons and $%
f-$spin fluctuations. This is consistent with recent results from field
theory\cite{tsvelik} and exact diagonalization\cite{tsutsui}.

We wish to thank S. Fujimoto for indicating ref. $[6]$ to us. This work was
supported by a grant from the Natural Sciences and Engineering research
Council (NSERC) of Canada and the Fonds pour la formation de Chercheurs et
d'Aide \`a la Recherche (FCAR) of the Qu\'ebec government.

\figure FIG. 1. The electron momentum distribution $n_c(k)$ for $\rho
=0.75,0.5$ and $0.25$ at $J_K=10$ and $J_H=0.5$.

\figure FIG. 2. The magnetic structure factor $S_f(k)$ for $\rho =0.75,0.5$
and $0.25$ at $J_K=10$ and $J_H=0.5$.

\figure FIG. 3. $n_c(k)$ for various values of $J_K$ at $\rho =0.5$ and $%
J_H=0.5.$

\figure FIG. 4. $S_f(k)$ for various values of $J_K$ at $\rho =0.5$ and $%
J_H=0.5.$

\figure FIG. 5. $S_f(k)$ for $J_K=1.25,1.5,1.75,2$ and $2.5$ at $\rho =0.75$
and $J_H=0$.

\figure FIG. 6. $n_c(k)$ for $J_K=1.5,2$ and $2.5$ at $\rho =0.75$ and $%
J_H=0 $.

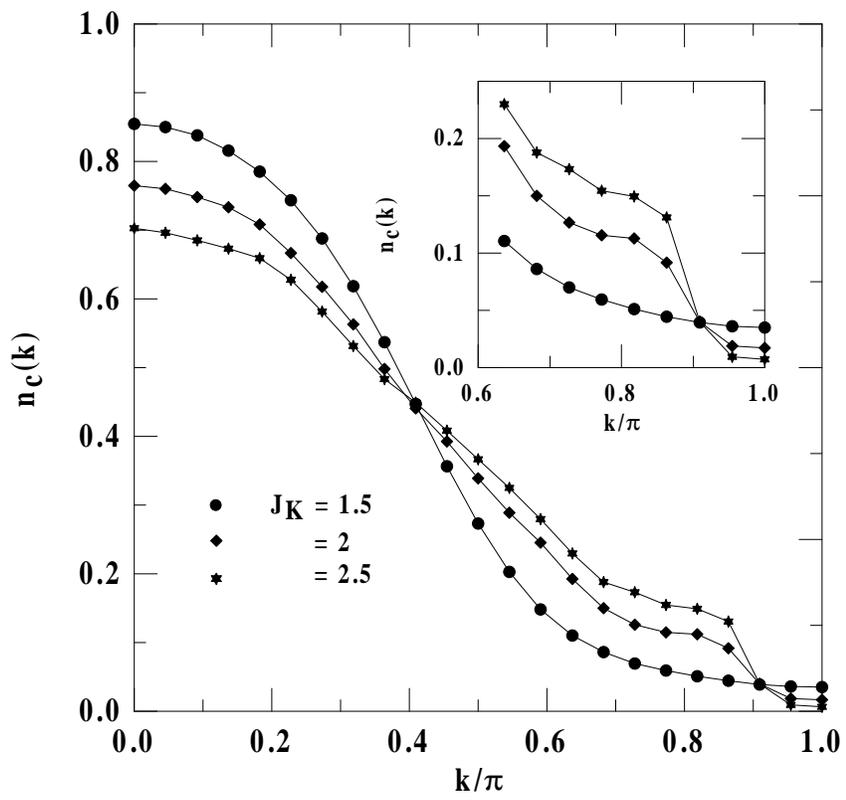
\end{document}